\documentclass[a4paper,UKenglish,numberwithinsect]{lipics-v2018}
 
\usepackage{microtype}
\usepackage{color}
\usepackage{booktabs}
\usepackage{algo}
\usepackage{thmtools, thm-restate}
\usepackage{cite}
\usepackage{graphicx}
\usepackage{xspace}
\usepackage[final]{changes}
\nolinenumbers

\newcommand{\eat}[1]{}
\newcommand{\set}[1]{\{#1\}}
\newcommand{\bag}[1]{\{\!\!\{#1\}\!\!\}}
\newcommand{\tpl}[1]{\bar #1}
\newcommand{\Qif}{\leftarrow}

\newcommand{\D}{{\mathcal D}}
\newcommand{\R}{{\mathcal R}}
\newcommand{\Q}{{\mathcal Q}}

\renewcommand{\S}{{\mathcal S}}

\newcommand{\ic}{{\mathcal I}}
\newcommand{\trans}{{\mathcal T}}
\newcommand{\possible}{{\mathit{Poss}}}
\newcommand{\canappend}[2]{\to_{#1,#2}}
\newcommand{\Canappend}[2]{\Rightarrow_{#1,#2}}
\newcommand{\pdc}{\mathrm{DCSAT}}
\newcommand{\MSEP}{\mathrm{MSEP}}
\newcommand{\KSEP}{\mathrm{KSEP}}
\newcommand{\comp}{\mathbin{\theta}}
\newcommand{\COUNT}{{\mathsf{count}}}
\newcommand{\COUNTD}{{\mathsf{cntd}}}
\newcommand{\MIN}{{\mathsf{min}}}
\newcommand{\MAX}{{\mathsf{max}}}
\newcommand{\SUM}{{\mathsf{sum}}}

\theoremstyle{plain}
\newtheorem{problem}{Problem}
\newtheorem{myremark}{Remark}

\newcommand{\ssty}[1]{{\sf{#1}}}
\newcommand{\tin}{{\mathit{in}}}
\newcommand{\tout}{{\mathit{out}}}
\newcommand{\Tin}{\trans_\tin}
\newcommand{\Tout}{\trans_\tout}
\newcommand{\fdalgname}{MSEPfd}
\newcommand{\fdalg}{\qproc{\fdalgname}\xspace}
\newcommand{\indalgname}{MSEPind}

\newcommand{\myheading}[1]{{\noindent\sffamily \bfseries #1.}}
\newcommand{\myheadingnl}[1]{\mbox{}\\\myheading{#1}}

\authorrunning{S. Cohen and A. Zohar}

\begin{document}
\title{Database Perspectives on Blockchains}

\author{Sara Cohen, Aviv Zohar}{The Rachel and Selim Benin School of Computer Science and Engineering\\
	The Hebrew University of Jerusalem\\
	Jerusalem, Israel}{}{}{}


\maketitle

\begin{abstract}
Modern 	blockchain systems are a fresh look at the paradigm of distributed computing, applied under new assumptions of large-scale open public networks.  They can be used to store and share information without a trusted central party. There has been much theoretical and practical effort to develop blockchain systems for a myriad of uses, ranging from cryptocurrencies to identity control, supply chain management, clearing and settlement, and more. However, none of this work has directly studied the many fundamental database-related issues that arise when  \replaced{using blockchains as the underlying infrastructure to store and manage data}{appending transactions to a blockchain database}.

	The key difference between \added{using} blockchains \replaced{to store data}{databases} and centrally controlled databases (of any type) is that transactions are accepted to a blockchain via a consensus mechanism, and not by the controlling central party. Hence, once a user has issued a transaction, she cannot be certain if it will be accepted. Moreover, a yet unaccepted transaction cannot be retracted by the user, and may be appended to the blockchain at any point in the future. This causes difficulties as the user may wish to reissue a transaction, if it was not accepted. Yet this data may then become appended twice to the blockchain.
		
	In this paper, we \added{present a database perspective on blockchains by} introducing formal theoretical foundations  for blockchains as \replaced{a data storage layer}{databases} that underlies a database. The main issue that we tackle is the uncertainty in transaction appending that is a direct result of the consensus mechanism. We  study two flavors of the transaction appending problem. First, we consider  the complexity of determining whether it is possible for a denial constraint to be contradicted, given the current state of the blockchain, pending transactions, and integrity constraints on blockchain data. Second, we determine the complexity of generating transactions that are (or are not) mutually consistent with given subsets of pending transactions. The ability to solve these problems is critical in ensuring that users can issue transactions that are consistent with their intentions. Finally, we chart directions for future work by discussing some of the new and important research challenges that arise when \replaced{data}{databases} are \added{stored using} blockchains.
	
 \end{abstract}


\section{Introduction}
Blockchains \deleted{databases} are becoming increasingly popular as a foundation for systems that  share information in a trustworthy way, but without a trusted central party. Among other uses, they form the basis for permissionless open systems in which participation is \deleted{completely} unregulated. 

Bitcoin,\footnote{We use the standard uppercase Bitcoin when referring to the system, and the lowercase bitcoin when referring to units of the currency.} created by Satoshi Nakamoto~\cite{nakamoto2008bitcoin},  is perhaps the prominent example of a permissionless blockchain system. It allows anyone to hold and transfer funds. To do so, rather than having a central server on which balances are kept, it decentralizes the management of funds between the many nodes in the Bitcoin network. Each node holds a copy of all transactions made within the blockchain. Bitcoin's appearance inspired the use of similar techniques to synchronize other forms of information in other systems, including many similar cryptocurrencies, but also brought a resurgence of interest in decentralizing existing systems, in the enterprise setting. 

In this paper we aim to explore some of the foundations of \replaced{blockchains as a storage level}{blockchain databases} in a manner that generalizes across different  protocols and consensus mechanisms. Conceptually, blockchain systems consist of a consensus layer that coordinates data synchronization between nodes (often in adversarial settings like Byzantine Fault Tolerance), and the data consistency layer that determines what are the legal states of the data and how it may evolve. In this paper we focus on the data consistency layer, and on the ramifications of the consensus layer on the possible states of the data.

\myheadingnl{Differences from classical database systems}
A Blockchain is typically an append-only \replaced{data structure}{database}, which is essentially a serialized record of accepted entries. Each individual block is an ordered batch of such updates that were committed together. Updates are  added through a consensus mechanism in which all participants  continually construct new blocks and accept them into the chain. 

This interaction is quite different from traditional database systems. While traditional systems may employ sharding or distribute the load of processing the data, they are still essentially keeping information under the control of \added{a {\em single}, possibly distributed, software system, which can be considered to be a} centralized entity. In blockchains, the main assumption is that participants are separate entities that are not controlled by the same administrator, and that records are accepted by consensus mechanisms.

Centralized entities are more efficient than blockchain systems, but in exchange are highly susceptible to the failure or corruption of that entity.  When only a single entity is in charge of record keeping, users may have their records erased, or altered without their permission. In currency systems, such interventions amount to freezing assets, or the seizure of funds without permission from their owner. Mainly for this reason, cryptocurrencies are implemented in a decentralized manner.

While updates to the blockchain become irreversible once committed (at least with very high probability), there is no guarantee that a certain update will indeed be accepted.\footnote{Some blockchain systems have guaranteed irreversibility, while others, like Bitcoin, do so probabilistically with probability approaching one as time proceeds.} 
Rules on the consistency of the database and the validity of transactions, that are applied by every individual node, effectively govern what enters the blockchain, in addition to external considerations that may affect nodes (such as fees paid to include certain transactions). 

Users that wish to add a record to the blockchain must create (and possibly sign) \replaced{an appropriate message.}{a message requesting this addition.} The message is \deleted{then }sent to the nodes that participate in the consensus mechanism for consideration. Thus, even an old record that was created and \deleted{similarly }propagated to other nodes by its original creator may be suddenly added to the blockchain. Users thus have a high level of uncertainty regarding transactions that have yet to be accepted. 

\myheadingnl{A motivating example}
As a simple motivating example, we will consider a Bitcoin exchange, that regularly accepts and issues payments in bitcoins in return for dollars or euros. Each payment that is issued \deleted{in bitcoins }by the exchange requires a miner's fee that is paid out to the node that approves the transaction and includes it inside a Bitcoin block. Fees form a part of the transaction message, and as such must be specified when the transaction is prepared.

Once the exchange broadcasts the transaction to the network, it must hope that it will be included in the blockchain. Unfortunately, since fees in the network can often fluctuate (depending on competition for limited space in Bitcoin blocks) it is quite possible for transactions not to be accepted by  the network, or to get delayed for extended periods. In this case, the exchange would still be interested in issuing the transaction, perhaps with an increased fee the second time around. Alas, once two transaction messages are broadcast to the network, it is again quite possible that both the new and the old message will be included in the blockchain. Once signed, a transaction can be rebroadcast to the network by anyone, and is in fact often stored by nodes, with the hope of later inclusion in the blockchain.  The unfortunate consequence of two such transactions making it to the blockchain is that the exchange would then pay its customer twice the amount it had intended. 

The design of Bitcoin transactions does include a remedy for this situation: two transactions can be made to conflict in such a way as to rule out their co-existence in the blockchain. Careful practices by the exchange would require that reissued transactions be set up in such a manner. In Section~\ref{sec:overview} we go into more details and explain the structure of Bitcoin transactions and the validity rules that introduce such relations between them. 

It is interesting to note that the problem we describe above is not merely hypothetical, but was actually used to attack exchanges in the past. Bitcoin transactions used to be somewhat malleable: one could change the contents of transactions in \deleted{certain }ways that would preserve their effects and maintain the validity of all cryptographic signatures. Attackers who  withdrew funds from exchanges then rebroadcasted a changed version of their withdrawal transaction which was later accepted into the blockchain. Unfortunately, the exchanges did not notice, and reissued the payments. This resulted in \deleted{some }withdrawals being issued more than once~\cite{decker2014bitcoin}.

While this example is specific to Bitcoin, it points to a deeper issue that arises when using a blockchain \replaced{to store data.}{ database.} Uncertainty as to which transactions will be committed may lead to different possible worlds. It is critical to be able to ascertain that there are no possibilities for undesirable outcomes. 
 
\myheadingnl{Contributions}
 Our main contributions can be summarized as follows. 
 \begin{itemize}
 	\item We present the first abstract model of \added{databases using blockchains as a storage layer, called} {\em blockchain databases}. Our model captures the main properties of such databases, while being independent of the specific protocol and implementation \added{of the blockchain}. \added{Since blockchains are already the underlying storage layer for cryptocurrencies, and are being extensively explored for other data storage scenarios, such a modelling is a first step towards studying the ramifications of using a blockchain as for storage in a database.}
 	\item We study the problem of determining whether a blockchain database can reach an undesirable state and provide complete complexity results. (A formal notion of desirability of outcomes is part of our model.)
 	\item We study a complementary problem of automatically generating transactions for the database. Here we aim to avoid undesirable situations by creating structures of dependencies and contradictions that will preclude undesirable states. We provide both algorithms and an undecidability result. 
 	\item Finally, we chart directions for future work by discussing some of the new and important research challenges that arise when databases are \added{based on} blockchains.
 	 \end{itemize}

Our paper is organized as follows. In Section~\ref{sec:overview}, we review Bitcoin's blockchain and structure of transactions, to ground our work in an important real-world implementation. Next, in Section~\ref{sec:related}, we review related work, pertaining to blockchains and to databases. 
 We present the formal framework in Section~\ref{sec:definitions} and define problems of interest in Section~\ref{sec:problems}. We study the denial constraint satisfaction problem in Section~\ref{sec:denial:constraints} and the problem of transaction generation in Section~\ref{sec:generating}.  We discuss important future research directions in Section~\ref{sec:open} and in Section~\ref{sec:conclusion} we conclude.  Due to space limitations, all proofs are deferred to the appendix.


\section{Bitcoin's blockchain and the structure of transactions} \label{sec:overview}
While the paper presents results that generalize to any blockchain database, due to its popularity, we use Bitcoin as our running example throughout the paper. In this section, we provide a brief overview of the Bitcoin protocol and its specific use of the blockchain.  The reader is referred to other, more-detailed explanations~\cite{zohar2015bitcoin} for more information, \added{and to \cite{IttaiDahlia} for a discussion of blockchain consensus protocols from the viewpoint of foundations of distributed computing}. \added{The goal of } this overview is to ground the discussion in concrete terms and relate the results to a real-world system.

The Bitcoin network is comprised of a set of nodes that together form a P2P network. Nodes in the network authorize Bitcoin transactions and include them in batches called blocks. Each block is a collection of transactions that transfer the bitcoin currency between different  addresses. Addresses  are effectively public keys that are associated with the money being transfered to denote ownership.

Each block additionally contains a cryptographic hash of a predecessor block. Blocks are thus arranged in a chain. The first block that was created at the inception of the system is known as the Genesis Block. The act of creation of blocks requires a difficult computation (proof-of-work). It rewards the creator of the block with bitcoins collected as fees from each transaction,  as well as bitcoins that are minted at a predetermined rate. The act of block creation is known as {\em mining,} and nodes that choose to invest the computational effort to try and create blocks are referred to as {\em miners}. 

As blocks can be created concurrently, slightly different versions of the chain may be held by different nodes. 
Miners reach consensus on the blockchain by selecting to adopt the chain with the most accumulated proof-of-work. They quickly distribute newly found blocks to one another as well as transaction messages that are meant for inclusion in blocks via a gossip protocol. Conflicting transactions, or blocks that are off the main chain, are not propagated and are immediately discarded. 

Each transaction in Bitcoin is constructed as a transfer of funds from inputs to outputs (many-to-many). Outputs are essentially an association between an amount of bitcoins and a script that specifies how this money is to be claimed. The typical script in Bitcoin requires that the spender present a valid cryptographic signature in order to spend funds, but other scripts are also possible, e.g., those requiring a preimage to a cryptographic hash to free funds, or perhaps several signatures matching different public keys. 
The inputs of each transaction essentially point to previously accepted outputs, and provide the necessary response to the challenge posed by the output scripts  that are needed to spend the money. Funds are then taken from the inputs and redistributed to the output of every transaction. 

\begin{figure}
	\centering\includegraphics[scale=0.4]{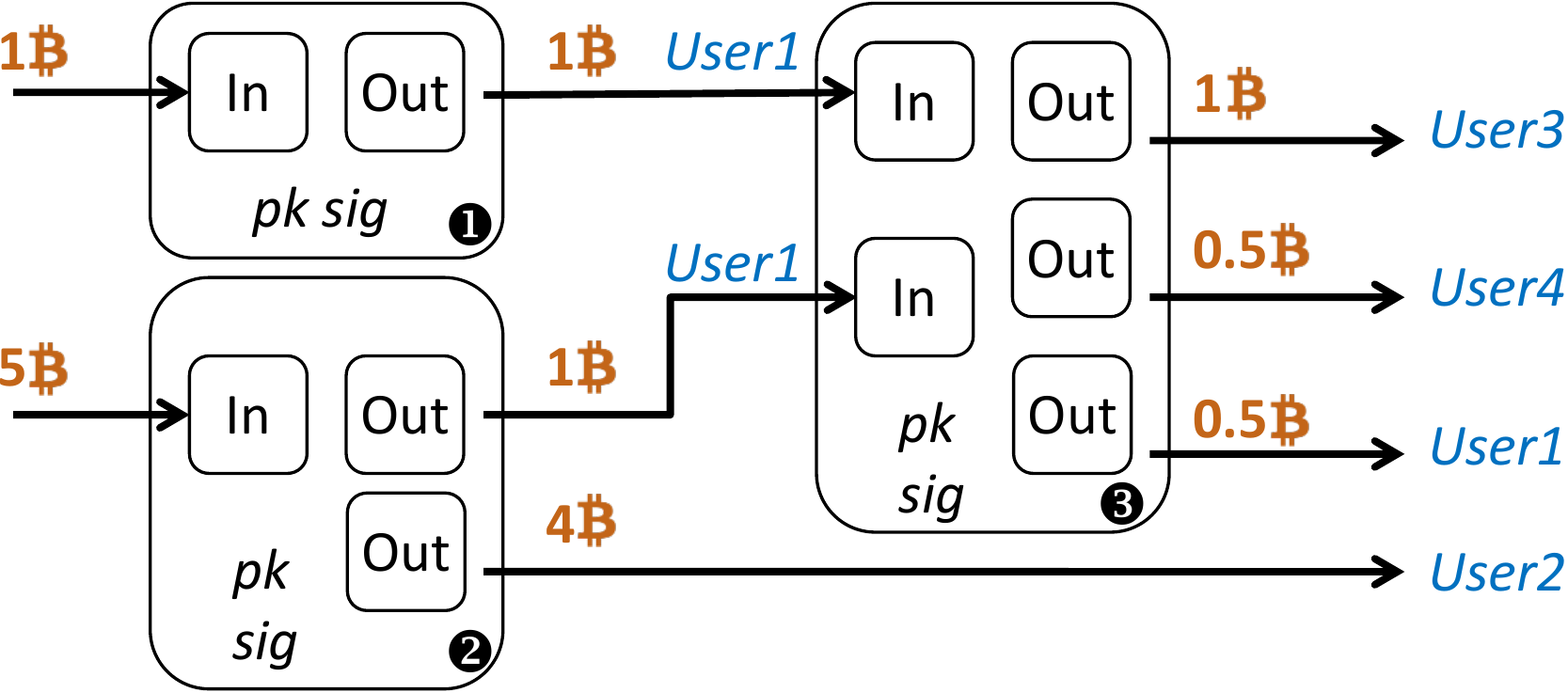}
	\caption{Structure of some Bitcoin transactions.\label{fig:bitcoin:diagram}}
\end{figure}

A Bitcoin transaction is considered to fully spend all inputs that it uses. Two transactions that share even a single input are thus considered conflicting transactions and cannot be accepted into the blockchain together. Transactions that spend the outputs of previous ones are dependent upon these parent transactions and cannot be included without first including the parent transactions. 
%
%
Figure~\ref{fig:bitcoin:diagram} shows an example of three Bitcoin transactions, along with their inputs and outputs.

\section{Related Work} \label{sec:related}

This paper touches on two  areas, namely {\em (1)\/}~blockchains (and bitcoin, in particular) and {\em (2)\/} uncertain and inconsistent databases. We discuss related work in each of these areas next.

\myheadingnl{Blockchains and Bitcoin}
There has been significant research on blockchains in general, and on Bitcoin in particular. For example, recent work has studied susceptibility of Bitcoin to selfish mining attacks (in which users selectively delay publishing data)~\cite{DBLP:conf/fc/EyalS14,Sapirshtein2017}, protocols to incentivize data propagation~\cite{Babaioff:2012:BRB:2229012.2229022}, to ensure correct incentives in mining pools~\cite{rosenfeld2011analysis, lewenberg2015bitcoin}, and modifications to the consensus protocol that will enable better scalability~\cite{DBLP:conf/nsdi/EyalGSR16,DBLP:journals/iacr/GiladHMVZ17, sompolinsky2016spectre, sompolinsky2015secure}. 

The structure of transactions in Bitcoin allows for de-anonymization and analysis~\cite{ron2013quantitative,reid2013analysis}, which lead to several works that improved the privacy of cryptocurrencies~\cite{miers2013zerocoin, sasson2014zerocash, danezis2013pinocchio, DBLP:conf/sp/KosbaMSWP16}. The security of the protocol to double-spending attacks was widely researched~\cite{sompolinsky2016bitcoin, garay2015bitcoin, karame2012double}. Other works have dealt with the susceptibility of Bitcoin to networking attacks~\cite{heilman2015eclipse, apostolaki2016hijacking}. However, no work to date has taken a higher level view of a blockchain \replaced{as a storage layer,}{ database,} in which the implementation details of the mechanisms are abstracted away, and the fundamental querying problems are given full focus. 

\myheadingnl{Uncertain and Inconsistent Databases}
Conceptually, a  \replaced{a database using blockchains as a storage layer,}{  blockchain database,} consists of a current state (i.e., a set of relations, stored in the blockchain), integrity constraints that must hold in every state, and a set of append-only transactions that have been issued by users, but have not been incorporated (yet) into the blockchain. Thus, a blockchain database is a succinct representation of a  large number of possible worlds. Each possible world is a consistent set of relations, including the current state, and perhaps the results of additional  transactions from among those issued.  Possible worlds are not necessarily {\em maximal\/}, i.e., they may not contain some of the issued transactions, even though these transactions can be included while preserving consistency w.r.t.\ the integrity constraints. 

Previous work considered probabilistic databases~\cite{dalvi2007efficient,cavallo1987theory,sen2007representing} and inconsistent databases~\cite{DBLP:conf/pods/ArenasBC99,DBLP:journals/amai/StaworkoCM12,DBLP:journals/sigmod/Bertossi06,DBLP:conf/dagstuhl/BertossiC03,Chomicki:2004:CCQ:1031171.1031254,DBLP:journals/jcss/FuxmanM07,DBLP:journals/dke/MarileoB10,DBLP:journals/pvldb/KolaitisPT13}, both of which represent many possible worlds in a succinct manner. Querying is a challenge in both settings, as in the former setting, answers must be returned with their associated probabilities, and in the latter, only certain answers should be returned.

While a blockchain database is uncertain (as one cannot be sure which transactions will be accepted), it differs  from a probabilistic database. Notably, it is not clear how to realistically associate probabilities with transactions, as miners can decide which transactions to try to include in a block, based on many   considerations.
The results in this paper are somewhat in the spirit of previous work on inconsistent databases. However, the setting is significantly different, as are the problems studied, and thus, previous results cannot be used. First and foremost, a blockchain database is always consistent. Moreover, while repairs (for inconsistent databases) are typically maximally close to the given data, in some sense, possible worlds for blockchains need not be maximal in any sense. Hence, for example, conjunctive query answering is not a key problem for blockchain databases (see Section~\ref{sec:problems} for more details). 


\section{Formal Framework} \label{sec:definitions}

A {\em relation\/} $R(A_1,\ldots,A_n)$ is associated with a set of ground tuples of  arity $n$.\footnote{By abuse of notation we use $R$ denote both the schema of a relation, and its contents.} Given a set of relations $\R$, an {\em insert transaction\/} (or simply, {\em transaction\/}, for short) for $\R$ is a set $T$ of ground tuples for (some of) the relations in $\R$. Intuitively, $T$ is a set of tuples that we would like to insert into the relations of $\R$. We consider only insert transactions, as blockchain databases are append-only databases.

We consider three types of integrity constraints, namely, {\em key constraints\/}, {\em functional dependencies\/} and {\em inclusion dependencies\/}.  Recall that key constraints are a special case of functional dependencies. For completeness, we define these notions in the appendix. We define satisfaction of a set of integrity constraints $\ic$ by a set of relations $\R$ in the standard fashion, and denote this by $\R\models \ic$.

\begin{example}\label{example:bitcoin:db}
Consider a database containing a simplified version of bitcoin data, with two relations describing transaction outputs and inputs: 
\begin{align*}
 &\ssty{TxOutput}\ssty{(\underline{txId, ser}, pk, amount)} 
&\ssty{TxInput}\ssty{(\underline{prevTxId, prevSer}, pk, amount, newTxId, sig)}. 
\end{align*}
In \ssty{TxOutput}, attribute \ssty{txId} is a transaction identifier, \ssty{ser} is the serial number of the particular transaction output (of which there may be several),  \ssty{pk} is the public key of the entity receiving this particular output and \ssty{amount} is the number of bitcoins in the output. In \ssty{TxInput}, the values \ssty{prevTxId,  prevSer, pk, amount} indicate the transaction number, serial number, public key and amount of the input being  consumed. Attribute \ssty{newTxId} is the new transaction identifier, and  \ssty{sig} is a cryptographic signature corresponding to the public key of the entity  whose output is being consumed. 

The keys of both relations are underlined. We consider the inclusion dependencies 
\begin{align*}
\ssty{TxInput[prevTxId,prevSer, pk, amount]}&\subseteq \ssty{TxOutput[txId,ser, pk, amount]},\\
\ssty{TxInput[newTxId]}&\subseteq \ssty{TxOutput[txId]}
\end{align*}
i.e., every input consumed was created as the output of some transaction, and  every new transaction has outputs.  \qed
\end{example}

We now formally define the notion of a {\em blockchain database}.
A {\em blockchain database\/} $\D$ is a triple $(\R, \ic, \trans)$, where 
\begin{itemize}
\item $\R$ is a set of relations, called the {\em current state,\/}
\item $\ic$ is a set of integrity constraints, such that $\R\models \ic$,  
\item $\trans = \set{T_1,\ldots,T_k}$ is a finite set of transactions for~$\R$, each of which is a set of tuples.
\end{itemize}

\begin{figure}[t]
		\begin{tabular}[t]{l|llllll}
		\multicolumn{7}{c}{{\ssty{TxInput}}}\\ \toprule 
		 &\ssty{prevTxId} & \ssty{prevSer} & \ssty{pk} & \ssty{amount} & \ssty{newTxId} & \ssty{sig}  \\\midrule
		$\R$ & 1 &  1 & U1Pk & 1  & 3 & U1Sig \\  
		& 2 &  1 & U1Pk & 1  & 3 & U1Sig \\  \midrule
		$T_1$ & 2 &  2 & U2Pk & 4  & 4 & U2Sig \\ \midrule
		$T_2$ & 4 &  1 & U2Pk & 3  & 5 & U2Sig \\ \midrule
		$T_3$ & 3 &  3 & U1Pk & 0.5  & 6 & U1Sig \\ \midrule
		$T_4$ & 6 & 1  & U4Pk & 0.5 & 7 & U4Sig \\
		 & 5 & 1  & U4Pk & 3 & 7 & U4Sig \\\midrule
		$T_5$ & 2 & 2 & U2Pk & 4 & 8 & U2Sig \\
		\bottomrule 	\multicolumn{7}{c}{} \\ \multicolumn{7}{c}{} \\
	\multicolumn{7}{c}{\includegraphics[scale=0.35]{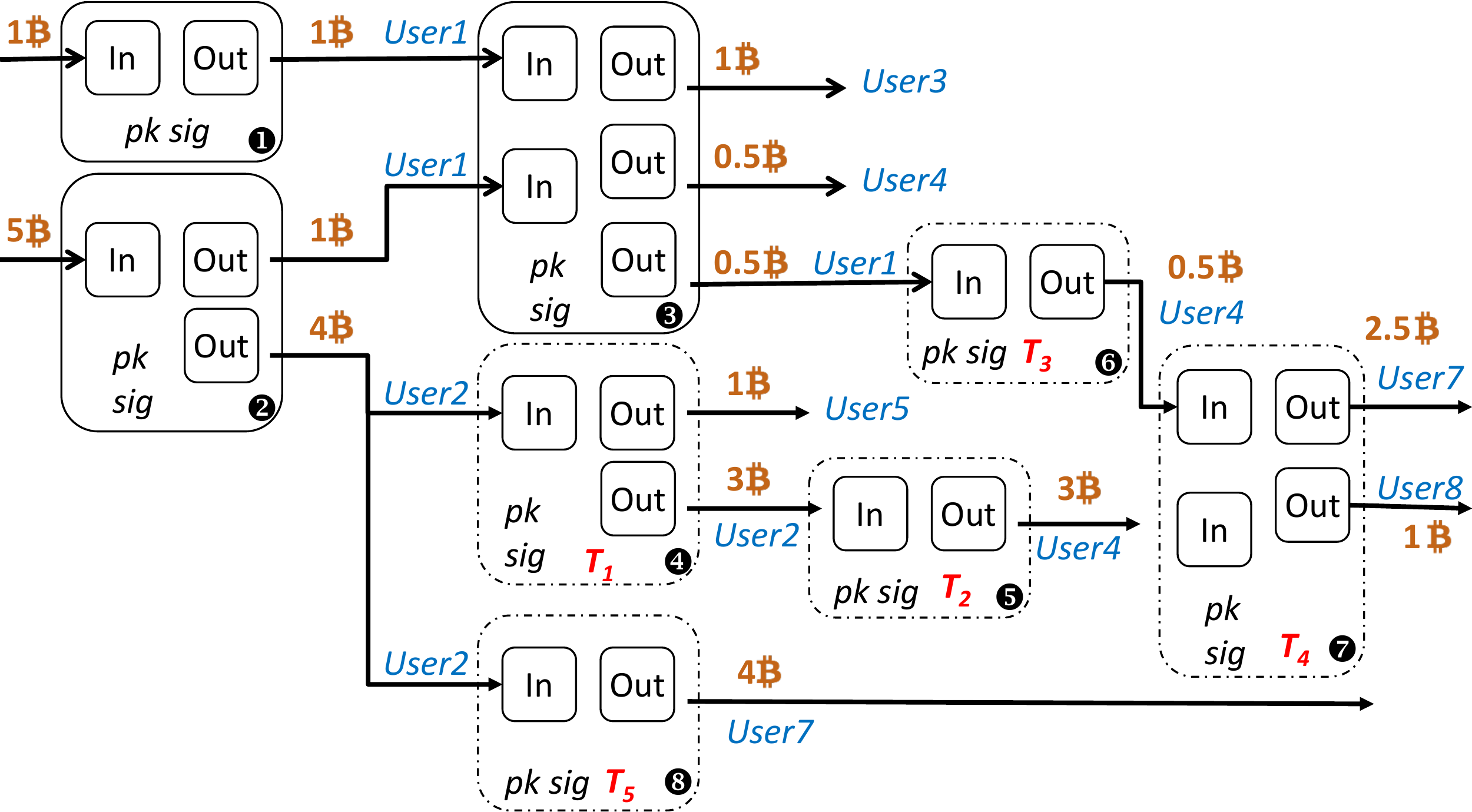}}
	\end{tabular}\hfill \ \ 
	\begin{tabular}[t]{l|llll}
		\multicolumn{5}{c}{{\ssty{TxOutput}}}\\ \toprule 
		& \ssty{txId} &  \ssty{ser} & \ssty{pk} & \ssty{amount} \\\midrule
	$\R$	& 1 &  1 & U1Pk & 1  \\  
		& 2 &  1 & U1Pk & 1  \\  
		& 2 &  2 & U2Pk & 4  \\  
		& 3 &  1 & U3Pk & 1 \\
		& 3 &  2 & U4Pk & 0.5 \\
		& 3 &  3 & U1Pk & 0.5 \\ \midrule
  $T_1$ & 4 &  1 & U5Pk & 1 \\
  & 4 &  2 & U2Pk & 3 \\ \midrule
  $T_2$ & 5 &  1 & U4Pk & 3 \\\midrule
  $T_3$ & 6 & 1  & U4Pk & 0.5 \\\midrule
  $T_4$ & 7 & 1 & U7Pk & 2.5 \\
     & 7 & 2 & U8Pk & 1 \\\midrule
    $T_5$ & 8 & 1 & U7Pk & 4 \\ \bottomrule  
	\end{tabular} \\
	\caption{Running example of a simplified Bitcoin blockchain database. \label{fig:running:example}}
\end{figure}

\begin{example}
Figure~\ref{fig:running:example} contains an extension of the diagram from Figure~\ref{fig:bitcoin:diagram} (depicting financial transactions) but now also includes transactions that have been issued and not yet been accepted into the blockchain (blocks with a dotted outline). Figure~\ref{fig:running:example} also contains an instance of a blockchain database, with the schema from Example~\ref{example:bitcoin:db}, reflecting the contents of the diagram.
In each table, the first column indicates whether the tuples belong to the current state $\R$, or to a transaction $T_i$. \qed
\end{example}

The transactions in $\trans$  may be appended to the database. 
 However, it is also possible that they will not be appended, e.g., because they are not mutually consistent with $\ic$, because of network problems, or simply due to lack of financial incentive.\footnote{In the real world, transactions are also associated with a fee that is  accrued when it is added to the current state.}

We define the {\em can-append relationship\/} $\R\canappend \trans\ic \R'$ for $\D$, as follows. We write $\R\canappend \trans\ic  \R'$ if $\R' = \R$ or there is some $T\in \trans$ such that $\R' = \R \cup T$ and $\R'\models \ic$. We denote the transitive closure of this relationship as $\R\Canappend \trans\ic \R'$.  Intuitively, this relationship defines a new instance of relations that can be derived by incrementally adding transactions from $\trans$, while preserving all integrity constraints. 
We say that $\R'$ is a {\em possible world\/} for $\D$ if $\R\Canappend \trans\ic \R'$, and denote the set of all possible worlds of $\D$ by $\possible(\D)$. Possible worlds can be efficiently recognized.\footnote{\added{In some systems it is possible that $\trans$ is not fully known, as transactions are issued by multiple users concurrently. However, often transactions that have been issued are propagated among all users.}} 

\begin{restatable}{proposition}{propPTIME}\label{prop:possible:recognize}
Let $\D$ be a blockchain database and let $\R'$ be a set of relations. It is possible to determine if $\R'\in \possible(\D)$ in PTIME. 
\end{restatable}

\added{We note that all tractability results in this paper show polynomiality in the size of the underlying blockchain and pending transactions. In practice, it is often possible to employ indexing in order to iterate only over a small portion of the blockchain. We leave such practical optimizations for future work.}

\newcommand{\propPTIMEProof}{
\begin{proof}
Let $\D = (\R,\ic,\trans)$ and let $\R'$ be a set of relations.
Let  $\trans'\subseteq \trans$ be the set of all transactions $T\in \trans$ such that $T$ is contained in $\R'$. 

To check if $\R'\in \possible(\D)$,  first determine if $\R'$ contains any contradictions to the key and functional dependencies in $\ic$. If so, clearly $\R'\not\in\possible(\D)$. Next, check if $\R\cup \trans' = \R'$. If not, again, $\R'$ cannot be a possible world. 
Finally, iteratively choose a transaction $T\in \trans'$ that can be added to $\R$ without contradicting any inclusion dependencies, and remove $T$ from $\trans'$. Continue this process until either we have derived $\R'$, in which case $\R'\in \possible(\D)$, or until no further progress can be made, in which case $\R'\not\in \possible(\D)$.
\end{proof}}

\begin{example}
Consider the blockchain database $\D = (\R, \ic, \set{T_1,T_2,T_3, T_4,T_5})$ depicted in Figure~\ref{fig:running:example}. 
Observe that transactions  $T_1$ and $T_5$ are not mutually consistent, as including both would contradict the key constraint on \ssty{TxInput} (intuitively, corresponding to a double-spend\-ing of the same money). It is interesting to note that one of the outputs of $T_2$ is given to User2, who is the same user creating $T_2$. This accurately reflects the manner in which transactions are performed in Bitcoin, as users return to their own wallet the remainder of the input not being sent to another user. Also note that the  transaction $T_4$ is dependent on $T_2$ and $T_3$ (due to the inclusion dependency), and $T_2$, in turn, is dependent on~$T_1$. 

Thus, $\possible(\D)$ contains $\R$, $\R\cup T_1$,  $\R \cup T_3$, $\R\cup T_1\cup T_3$, $\R\cup T_1 \cup T_2$, $\R\cup T_1\cup T_2\cup T_3$, $\R\cup T_1\cup T_2\cup T_3\cup T_4$, and $\R\cup T_5$.\qed
\end{example} 

\begin{myremark}
In our model of a blockchain database we purposely retain only the essentials that are required to study pivotal database issues. Thus, for example, we ignore the incentives, consensus mechanisms and confirmation policies used while constructing  blockchains. These are all orthogonal to our study, and also differ between various blockchain systems. We do not make explicit within $\R$ information about how tuples are distributed among the physical blockchain, as this is not of importance after data is accepted into the chain. (Transaction ordering is important, due to integrity constraints, while adding transactions, and this is specified in our model using the can-append relationship.) 
	
	We also do not  consider forks in the chain, which may create temporary inconsistent versions of the database, for two main reasons. First, whether or not such forks can exist, and how they are resolved is system dependent. (For example, Algorand~\cite{gilad2017algorand}, Solida~\cite{DBLP:journals/corr/AbrahamMNRS16} and ByzCoin~\cite{kogias2016enhancing} do not allow forking.) Second, typically forks are resolved within a few seconds and, from a theoretical standpoint, such uncertain data is not considered as included in the database (yet). 
\end{myremark}

\section{Problems of Interest}\label{sec:problems}
Over time, a blockchain database evolves, by the addition of transactions to the current state. However, at any given point in time, it is not possible to know for certain which pending transactions in $\trans$ will be permanently added to $\R$. One classical question that may be asked in this setting is as follows: 
{\em Given a query $q$ and a blockchain database $\D$, what are all certain answers to $q$ over~$\D$?}
Here certain answers are defined in the natural fashion (in the spirit of~\cite{Arenas:1999:CQA:303976.303983}), as tuples that will appear in the result over all possible worlds. In the setting of blockchain databases, it is not clear that this is of interest, e.g.,  for conjunctive queries, the set of certain answers is precisely the result of evaluating $q$ over $\R$. 

Looking at this from the opposite perspective, recall that users often issue many transactions, and during this process they do not know for certain which previous transactions will or will not be appended to the current state. Hence, such a user may wish to ensure that in all possible worlds, specific undesirable things will not happen, by answering the following question:
{\em Given a Boolean query $q$, does $q$ evaluate to false over all possible worlds of $\D$?} 
We call such a Boolean query, which we desire to remain unsatisfied, a {\em denial constraint}. Formally,  a denial constraint $q$ is satisfied by a blockchain database $\D$, denoted $\D\models\neg q$, if, for all $\R'\in \possible(\D)$, it holds that $q(\R') = \text{false}$. 

\begin{example}\label{example:denial:constraint}
Suppose Alice wishes to pay Bob one bitcoin (e.g., for a service Bob provided). Alice adds a transaction containing two tuples:
\begin{enumerate}
\item  a tuple for \ssty{TxInput}, with information about her bitcoin that will be consumed,\footnote{This, in turn, must be an output from some previous transaction in which Alice received this coin.} and 
\item a tuple for \ssty{TxOutput} with  the public key information of Bob. 
\end{enumerate}
After some time passes, Bob complains that he never received payment, i.e., that this transaction was not added to $\R$. Therefore, Alice sends a new transaction, again containing two tuples, to transfer money to Bob. Now, Alice may want to be certain that in all possible worlds, she has sent at most one bitcoin to Bob. 

Assuming AlicePK is the public key of Alice, and BobPK is the public key of Bob, she may use the denial constraint
\begin{align*}
 q_1() \Qif &\ssty{TxInput}(pt_1, ps_1, \text{\rm`AlicePK'}, 1, ntx_1, \text{\rm`AliceSig'}),
\ssty{TxOutput}(ntx_1, ns_1, \text{\rm`BobPK'}, 1),\\ 
&\ssty{TxInput}(pt_2, ps_2, \text{\rm`AlicePK'}, 1, ntx_2, \text{\rm`AliceSig'}),
\ssty{TxOutput}(ntx_2, ns_2, \text{\rm`BobPK'}, 1),\\ & ntx_1\neq ntx_2
\end{align*}
specifying that there are two different transactions in which Alice transferred money to Bob.

In practice, Alice may use this denial constraint over the hypothetical situation in which she has issued the second transaction (i.e., as a dry run, before actually issuing), in order to determine the safety of performing the second transaction. \qed
\end{example}

Being able to determine whether a denial constraint is satisfied by a blockchain database is a fundamental issue, that allows users to interact with the database with some degree of clarity as to the possible downsides of their transactions. Hence, in this paper we study the complexity of this problem. 
Section~\ref{sec:denial:constraints} presents a formal definition of the denial constraint satisfaction problem and studies the complexity of this  problem for different classes of queries, as well as for different types of integrity constraints. 

One way to avoid undesirable possible worlds, is by checking satisfaction of denial constraints, before issuing a new transaction. Another complementary method is to purposely generate a new transaction in such a fashion that will not allow it to be included in a possible world with specific other transactions. Intuitively, the problem of automatically deriving such new transactions can be described as follows:
{\em Given a blockchain database $(\R,\ic,\trans)$, generate a transaction $T$ that is inconsistent with a subset of transactions $\trans_1\subseteq \trans$, but is consistent with $\trans_2\subseteq \trans$. }
Note that we say that $T$ is inconsistent  with $\trans_1$ if there is no possible world containing both $T$ and any $T'\in \trans_1$, while $T$ is consistent with $\trans_2$ if there is a possible world containing $T$ and all of $\trans_2$. 

\begin{example}\label{example:generating}
As in Example~\ref{example:denial:constraint}, suppose that Alice wishes to pay Bob one bitcoin. Alices issues a transaction $T_1$ transferring money as before. Again,  time passes, and Bob complains that he never received payment. Now, Alice would like to generate a new transaction $T_2$ such that $T_2$ cannot appear together with $T_1$ in any possible world (to avoid double payment). In addition, Alice may desire that  inclusion of $T_2$ in a possible world should not interfere with some  other previous transactions which she issued. 

For example, suppose $T_1$ is the set of tuples
$\{\ssty{TxInput}(1037, 2, \text{\rm`AlicePK'}, 1, 5043, \text{\rm`AliceSig'}),$ $\ssty{TxOutput}(5043, 1, \text{\rm`BobPK'}, 1)\}$.
If all Alice desires is a new transaction inconsistent with $T_1$, the following transaction $T_2$ could be generated: 
$\{\ssty{TxInput}(1037, 2, \text{\rm`AlicePK'}, 1, 6074, \text{\rm`AliceSig'}),$ $\ssty{TxOutput}(6074, 1, \text{\rm`BobPK'}, 1)\}$.
Observe that $T_2$ and $T_1$ cannot both be in a possible world together, as this would contradict the key constraint over \ssty{TxInput}. \qed
\end{example}

The ability to automatically generate a new transaction, given a set of transactions with which it should be (in)consistent, is very useful when interacting with a blockchain database. This problem is  defined and studied in Section~\ref{sec:generating}, for different classes of integrity constraints. 

\section{Denial Constraint Satisfaction}\label{sec:denial:constraints}

We formally define the denial constraint satisfaction problem, discussed earlier. 

\begin{problem}[Denial Constraint Satisfaction]
Let $\D$ be a blockchain database and $q$ be a denial constraint. Determine whether $\D\models\neg q$.
\end{problem}
We use the measure of data complexity (i.e., we assume the denial constraint is of constant size, while the size of the database is unbounded), as otherwise, query evaluation over a standard database is already intractable. 

The complexity of the denial constraint satisfaction  problem varies greatly, depending on the language of the denial constraints, as well as the types of integrity constraints allowed. In the following we use $\Delta$ to denote a subset of $\set{key,fd,ind}$ (standing for key, functional dependency, inclusion dependency).  A blockchain database is {\em allowed by\/} $\Delta$, if it  only contains integrity constraints of types appearing in $\Delta$.  

Given a class of Boolean queries $\Q$, and a set  $\Delta$,
we use $\pdc(\Q,\Delta)$ to denote the denial constraint satisfaction problem for the given class of queries and types of integrity constraints. 
Formally,  $\pdc(\Q,\Delta)$ is in complexity class $\mathcal C$ if, 
for all $q\in \Q$ and 
for all blockchain databases $\D$ allowed by $\Delta$,
it is possible to determine whether $\D\models \neg q$ in time~$\mathcal C$. 

We note in passing that all hardness results for functional dependencies already hold when there are only key constraints, and that all hardness results for inclusion dependencies already hold even when there is a single inclusion dependency.

\myheadingnl{Classes of Queries.}
We consider several classes of  queries, defined next. A {\em conjunctive query\/} has the form 
$q() \Qif P, N, C$
where $P$ is a conjunction of positive relational atoms, $N$ is a conjunction of negated relational atoms and $C$ is a conjunction of comparisons of variables to one another or to constants (using $=, <, >$, or $\neq$). We assume that all queries are {\em safe\/}, i.e., every variable appears in a positive relational atom. A conjunctive query is {\em positive\/} if it does not contain any negated relational atoms. 

Given a set of relations $\R$, the query $q$ returns true if there is an {\em satisfying assignment\/} $h$ of the variables in the body of $q$ to constants in $\R$ such that {\em (1)}
 for all positive relational atoms $a$, $h(a)\in \R$, {\em (2)}
 for all negated relational atoms $a$, $h(a)\not\in\R$ and {\em (3)}
 $h(C)$ is satisfied.
  Otherwise, $q$ returns false. We use $\Q_c$ to denote the class of conjunctive queries and $\Q^+_c$ to denote the subclass containing only positive queries. 

An {\em aggregate function\/} $\alpha(\tpl x)$, where $\tpl x$ is an $n$-ary tuple of variables, is applied to a bag of $n$-ary tuples of values, and returns a single value.\footnote{We allow $n=0$ to accommodate the aggregate function $\COUNT$, when applied to bags of empty tuples.} A {\em aggregate query\/} has the form
$\left[q(\alpha(\tpl x)) \Qif P, N, C\right]\comp c$
where $\theta\in \set{=,<,>}$ is a comparison and $c$ is a constant. 

Given a set of relations  $\R$, let $H$ be the set of all satisfying assignments of the body of $q$ into $\R$. Let $B$ be the bag $\bag{h(\tpl x) \mid h\in H}$. Then, the query $q$ returns true if and only if $\alpha(B)\comp c$ is true. For the special case in which $B$ is empty, we 
consider $\alpha(B)\comp c$  to be false.\footnote{This choice of semantics seems to be closest to the spirit of SQL. However, one can also choose to consider  $\alpha(B)\comp c$  to be true in this case, or to determine the truth value based on $\comp$. Such changes in semantics will change some of the complexity results.} 
The syntax and semantics of the aggregate functions considered in this paper appear in the appendix.
We use $\Q_\alpha$ to denote the class of all aggregate queries using $\alpha$, and $\Q_\alpha^+$ to denote its positive subclass. 
Finally, we use $\Q_{\alpha,\comp}$ and $\Q_{\alpha,\comp}^+$ to denote the subclasses of  $\Q_{\alpha}$ and $\Q_{\alpha}^+$, respectively, in which the comparison operator is $\comp$. 


\begin{example}
The denial constraint $q_1$ in Example~\ref{example:denial:constraint} is a positive conjunctive query. We demonstrate additional denial constraints.
Suppose the relation \ssty{Trusted(\underline{pk})} contains a list of private keys of trustworthy individuals. 
Then $q_2$,  ensuring that all bitcoins from Alice are given to trusted individuals, is a conjunctive query, but is not positive. 
\begin{align*}
 q_2() \Qif &\ssty{TxInput}(pt, ps,  \text{\rm`AlcPK'}, a, ntx, \text{\rm`AlcSig'}),  \ssty{TxOutput}(ntx,s,pk,a'), \neg \ssty{Trusted}(pk)
\end{align*}
Denial constraint $q_3$ requires that Alice spent at most five bitcoins in total (as otherwise, it would return the value true)
\begin{align*}
 [q_3(\SUM(a)) \Qif &\ssty{TxInput}(t, s,  \text{\rm`AlcPK'}, a, nt, \text{\rm`AlcSig'})]  > 5 \,,
\end{align*}
while $q_4$ ensures that Alice participated in at most ten transactions in which bitcoins were given to Bob
\begin{align*}
 [q_4(\COUNTD(ntx)) \Qif & \ssty{TxInput}(pt, ps,  \text{\rm`AlcPK'}, a, ntx, \text{\rm`AlcSig'}),  
\ssty{TxOutput}(ntx,s,\text{\rm`BobPK'},a')]  > 10\,.
\end{align*}
\end{example}

\myheading{Complexity.}
Before studying the complexity of $\pdc(\Q,\Delta)$ for specific cases of $\Q$ and $\Delta$, we observe the following result that provides an upper bound on the complexity of the problems at hand. 

\begin{corollary}\label{cor:coNP}
$\pdc(\Q,\Delta)$ is in CoNP.
\end{corollary}

This result is a corollary of Proposition~\ref{prop:possible:recognize}. To see this, suppose  $q$ is a denial constraint and $\D$ is a blockchain database $(\R,\ic,\trans)$. Given any subset $\trans'\subseteq \trans$ and an instance $\R' =\R\cup \trans'$ it is possible to determine in PTIME whether $\R'$ is a possible world of $\D$ (Proposition~\ref{prop:possible:recognize}). Since we assume that $q$ is of constant size, it is possible to determine in PTIME whether $q$ evaluates to true over $\R'$. This provides us with the   upper bound.

The following theorem states the complexity of denial constraint satisfaction for different combinations of query classes and types of integrity constraints. We note that this theorem provides a full characterization for non-aggregate denial constraints, as the cases discussed clearly imply the complexity of the remaining cases, e.g.,  $\pdc(\Q_c, \set{key, ind})$ is CoNP-complete, since $\pdc(\Q^+_c, \set{key, ind})$ is CoNP-complete, and all problems are in CoNP by Corollary~\ref{cor:coNP}. 

\begin{restatable}{theorem}{thmComplexityConjunctive}\label{thm:complexity:conjunctive}
The denial constraint satisfaction problem has the following complexity:
\begin{enumerate}
\item  $\pdc(\Q_c, \set{key,fd})$  and $\pdc(\Q_c, \set{ind})$ are in PTIME.
\item $\pdc(\Q^+_c, \set{key, ind})$ is CoNP-complete. 
\end{enumerate}
\end{restatable}

\newcommand{\thmComplexityConjunctiveProof}{
\begin{proof}
We first show that $\pdc(\Q_c, \set{key,fd})$ is in PTIME.
Let $q$ be a denial constraint with $k$ positive relational atoms in its body and let $\D = (\R,\ic,\trans)$ be a blockchain database. Clearly, there is a possible world $\R'$ for which $q$ returns true only if there is one derived by a subset of transactions $\trans'$ of size at most $k$. To see this, observe that at most one transaction is needed to satisfy each positive relational atom. Note that if there were inclusion dependencies, this might not be sufficient, as one transaction might imply the inclusion of another. 

Now, for every subset of $\trans' \subseteq \trans$ of size $k$ or less, we 
\begin{itemize}
\item check if $\R\cup \trans'$ is a possible world {\em and\/}
\item evaluate $q$ on $\R\cup \trans'$, to determine if $q$ returns true.
\end{itemize}
If all $\R\cup \trans'$ return false, we can conclude that $\D\models \neg q$. This procedure is in PTIME, as $k$ is a constant. 

Next, we show that $\pdc(\Q_c, \set{ind})$ is in PTIME. Let $P$, $N$ and $C$ be the positive, negated atoms and comparisons in the body of $q$. We will consider every possible mapping of the variables in $q$ to constants in  $\R$ or $\trans$, and check if there is a possible world, for which this mapping produces the Boolean value true.  To be precise, for each mapping $h$ of the variables in $q$ to constants in $\R$ or $\trans$, such that $h(P)\subseteq \R\cup_{T\in \trans} T$, $h(N)\cap \R = \emptyset$ and $h(C)$ is satisfied, we
\begin{itemize}
\item remove all transactions from $\trans$ containing any atom from $h(N)$;
\item iteratively choose a transaction $T\in \trans$  that can be added to $\R$ without contradicting any inclusion dependencies, and remove $T$ from $\trans$. We continue this process until either $\trans = \emptyset$, or until no further progress can be made. Let $\R_h$ be the result. (Note that the result of this process is well defined, as we only consider inclusion dependencies.)
\item check if the $\R_h$ contains $h(P)$. If so, this is a counter-example to the denial constraint.
\end{itemize}
Since $q$ is of constant size, this process is polynomial, and clearly finds a counter-example if one exists. 

Finally, we show that  $\pdc(\Q^+_c, \set{key, ind})$ is CoNP-complete. Membership in CoNP follows from Corollary~\ref{cor:coNP}. We show hardness by a reduction from the complement of 3-SAT. Let $\phi_1\wedge \cdots \wedge \phi_n$ be a 3-SAT formula, where $\phi_i$ is a disjunction of variables or their negation from $x_1,\ldots,x_m$. We define a database with the relations Clause1($C$), Val($X,B$), Clause2($C$) and Sat($S$). Initially, all relations are empty. The attribute $X$ is a key in Val, and there is an inclusion constraint Clause2[$C$]$\subseteq$ Clause1[$C$]. 

 The set of transactions includes, for each literal $l$ in a disjunct $\phi_i$, the set $\{$Clause1($i$), Val($x,t$)$\}$ if $l$ is the positive literal $x$, or the set $\{$Clause1($i$), Val($x,f$)$\}$ if $l$ is the negative literal $\bar x$. In addition, there is a transaction $\{$Sat($t$), Clause2(1), $\ldots$, Clause2($n$)$\}$. Finally, the denial constraint is $q() \Qif $Sat$(t)$.
It is not difficult to see that there is a possible world in which the denial constraint evaluates to true if and only if the formula is satisfiable. Intuitively, the key constraint in Val ensures that the tuples in this relation correspond to a truth assignment. The inclusion dependency ensures that Clause2 will be populated only if Clause1 is populated, which, in turn, implies that the formula is satisfied. 
\end{proof}
}


We now consider query classes with aggregation. The complexity of the denial constraint satisfaction problem depends on the precise aggregate function considered. In the following theorem we provide a full characterization for the aggregate functions $\COUNT$, $\COUNTD$ (count distinct), $\SUM$ and $\MAX$. We note that the results for $\MAX$ can easily be used to determine the complexity for $\MIN$. The complexity results are also summarized in Table~\ref{table:complexity} in the appendix.

\begin{restatable}{theorem}{thmComplexityAgg}
Let $\alpha$ be an aggregation function. Then, 
\begin{enumerate}
\item  $\pdc(\Q_\MAX, \set{key,fd})$  is in PTIME.
\item $\pdc(\Q_{\alpha,<}, \set{key,fd})$, with $\alpha\in \{\COUNT, \COUNTD,$ $\SUM\}$, is in PTIME.
\item $\pdc(\Q^+_{\alpha,\comp}, \set{key})$, with $\alpha\in \{\COUNT, \COUNTD,$ $\SUM\}$, $\comp\in \set{>,=}$,
 is CoNP-complete.

\item  $\pdc(\Q^+_{\alpha,>}, \set{ind})$, with   $\alpha\in \{\COUNT,$ $\COUNTD, \SUM,$ $\MAX\}$, is in PTIME.
\item $\pdc(\Q^+_{\alpha,\comp}, \set{ind})$, with $\alpha\in \{\COUNT, \COUNTD, \SUM,$ $\MAX\}$, $\comp\in \set{<,=}$,
 is CoNP-complete.

\item $\pdc(\Q_{\alpha,>}, \set{ind})$, with $\alpha\in \set{\COUNT, \COUNTD, \SUM}$, is CoNP-complete.
\item $\pdc(\Q_{\MAX,>}, \set{ind})$ is in PTIME.
\item $\pdc(\Q^+_{\MAX} \set{key,ind})$ is CoNP-complete.
\end{enumerate}
\end{restatable}

\newcommand{\thmComplexityAggProof}{
\begin{proof}
We start by showing Claim~1 and~2. These results can be proven in the same manner as the proof that $\pdc(\Q_c, \set{key,fd})$ is in PTIME, in Theorem~\ref{thm:complexity:conjunctive}. Intuitively, the same proof (which considers all databases including at most $k$ transactions), is sufficient in this case, too, since if there is a database over which $q$ returns true, there is also such a ``small'' database.

We now consider  Claim~3.
We start by assuming that $\alpha = \COUNT$, $\comp$ is ``$>$'' and show hardness by a reduction from the complement of 3-SAT. Let $\phi_1\wedge \cdots \wedge \phi_n$ be a 3-SAT formula, where $\phi_i$ is a disjunction of variables or their negation from $x_1,\ldots,x_m$. We define a database with the relations Clause($C$) and Val($X,B$). Initially, all relations are empty. The attribute $X$ is a key in Val.

 The set of transactions includes, for each literal $l$ in a disjunct $\phi_i$, the set $\{$Clause($i$), Val($x,t$)$\}$ if $l$ is the positive literal $x$, or the set $\{$Clause($i$), Val($x,f$)$\}$ if $l$ is the negative literal $\bar x$. Finally, the denial constraint is $[q(\COUNT) \Qif $Clause$(i)] >  n-1$. It is not difficult to see that there is a possible world in which the denial constraint evaluates to true if and only if the formula is satisfiable. Intuitively, the key constraint in Val ensures that the tuples in this relation correspond to a truth assignment. The denial constraint  requires there to be at least $n$ satisfied clauses. Hence, the denial constraint is satisfied only if there is no possible world in which all clauses are satisfied, i.e., if the formula is unsatisfiable. 

For $\COUNTD$ and $\SUM$, can use the same construction, but with the denial constraints:
\begin{align*}
 [q(\COUNTD(i)) \Qif \text{Clause}(i)] & >  n-1 &
 [q(\SUM(i)) \Qif \text{Clause}(i)] & >  \frac{n(n+1)}{2}-1
\end{align*}
Finally, when $\comp$ is ``='' we use the same denial constraints, but add one to the number on the righthand-side and replace the comparison operator with ``=''.

We show Claim~4. Let $\R_{max}$ be the database derived by iteratively adding transactions from $\trans$ to $\R$ while satisfying all inclusion dependencies. (Note that $\R_{max}$ is unique as we only consider inclusion dependencies.) It is easy to see that for all $\R'\in\possible(\D)$ it holds that $\R\subseteq \R'\subseteq \R_{max}$. Hence, when the comparison operator is ``$>$'' it is sufficient to check whether $q$ returns true over $\R_{max}$. 

We show Claim~5 by a reduction from the complement of 3-SAT.  Let $\phi_1\wedge \cdots \wedge \phi_n$ be a 3-SAT formula, where $\phi_i$ is a disjunction of variables or their negation from $x_1,\ldots,x_m$. We define a database with the relations Clause1($C$), Val($X,B$), Clause2($C$), Sat($S$). Initially, all Clause1, Clause2 and Sat are empty. The relation Val has two rows $(0,t)$ and $(0,f)$. 
There is an inclusion constraint Clause2[$C$]$\subseteq$ Clause1[$C$]. 

 The set of transactions includes, for each literal $l$ in a disjunct $\phi_i$, the set $\{$Clause1($i$), Val($x,t$)$\}$ if $l$ is the positive literal $x$, or the set $\{$Clause1($i$), Val($x,f$)$\}$ if $l$ is the negative literal $\bar x$. In addition, there is a transaction $\{$Sat($t$), Clause2(1), $\ldots$, Clause2($n$)$\}$. Finally, we consider the following denial constraints (according to the value of $\alpha$):
\begin{align*}
 [q(\COUNT) &\Qif \text{Sat}(t), \text{Val}(x,b), \text{Val}(x,b'), b\neq b' ]  < 3 \\
[q(\COUNTD(x)) &\Qif \text{Sat}(t), \text{Val}(x,b),  \text{Val}(x,b'), b\neq b' ]  < 2 \\
[q(\SUM(x)) &\Qif \text{Sat}(t), \text{Val}(x,b), \text{Val}(x,b'), b\neq b' ]  < 1 \\
[q(\MAX(x)) &\Qif \text{Sat}(t), \text{Val}(x,b),  \text{Val}(x,b'), b\neq b' ]  < 1 
\end{align*}
(If $\comp$ is ``='' we replace ``$<$'' with ``='' and decrease the number of the righthand-side by one. )
It is not difficult to see that there is a possible world in which the denial constraint evaluates to true if and only if the formula is satisfiable. To see why, recall that an aggregate denial constraint can evaluate to true only if there is a satisfying assignment for its body. Thus, e.g., the first denial constraint above will return true only if there is a possible world in which Sat($t$) is included, and this world contains only one value for $x$ associated with two different $b$-s, i.e., precisely only $(0,t), (0,f)$. This will yield exactly two satisfying assignments for the body of the constraint.

We show Claim~6. We use a similar setting to Claim~5, but add an additional relation Truth($T$) containing two rows $(t), (f)$, as well as a transaction $\{$Val$(x,t)\}$ and $\{$Val$(x,f)\}$, for every variable $x$. 
Let $B(t,x,b,b')$ be the conjunction 
\[\text{Sat}(t), \text{Val}(x,b), \neg \text{Val}(x,b'), \text{Truth}(b').\]
We use  the following denial constraints (according to the value of $\alpha$):
\begin{align*}
 [q(\COUNT) &\Qif B(t,x,b,b') ]  > n-1 \\
[q(\COUNTD(x)) &\Qif B(t,x,b,b')]  >n-1 \\
[q(\SUM(x)) &\Qif B(t,x,b,b')]  > \frac{n(n+1)}{2}-1
\end{align*}
which return true when a possible world includes Sat($t$), and all variables have a single truth assignment. 

Claim~7 can be shown using essentially the same proof as that of $\pdc(\Q_c, \set{ind})$  in Claim~1 of Theorem~\ref{thm:complexity:conjunctive}. 
Finally, Claim~8 can be shown using essentially the same proof as that of Claim~2 in Theorem~\ref{thm:complexity:conjunctive}, using the denial constraint
\[ [q(\MAX(x)) \Qif \text{Sat}(x)] \comp t_{\comp}\]
where $t_=$ is $t$, $t_<$ is $ t+1$ and $t_>$ is $t-1$. Note that is is correct, as by definition, a denial constraint can only return true if there is a satisfying assignment for its body.
\end{proof}
}


\section{Generating Transactions} \label{sec:generating}
We formally define the problem of generating a transaction that is consistent with some given transactions, but not with others, as discussed in Section~\ref{sec:problems}.
Let $\D = (\R,\ic,\trans)$ be a blockchain database. Let $\Tin$ and $\Tout$ be  subsets of $\trans$. Let $T$ be a transaction and $\D' = (\R,\ic,\trans\cup \set{T})$.  

We say that $T$ is {\em mutually consistent\/} with $\Tin$ if there exists $\R'\in  \possible(\D')$, such that $(T\bigcup _{T'\in \Tin} T') \subseteq \R'$. We say that $T$ is {\em inconsistent\/} with $\Tout$ if for all $\R'\in \possible(\D')$, if $T\subseteq \R'$, then for all $T'\in \Tout$, it holds that $T'\not\subseteq \R'$.
Finally, we say that $T$ is {\em $(\Tin,\Tout)$-separating\/} for $\D$ if $T$ is both mutually consistent with $\Tin$ and inconsistent with $\Tout$. 

In general, there may be many $(\Tin,\Tout)$-separating transactions $T$, and such transactions may be of various sizes. We say that $T$ is {\em minimal\/} if no strict subset of $T$ is $(\Tin,\Tout)$-separating. We say that $T$ is {$k$-bounded\/} if $T$ contains at most $k$ tuples. 
We define the following two problems. In both, $\D=(\R,\ic,\trans)$ is a blockchain database, and $\Tin, \Tout\subseteq~\trans$.

\begin{problem}[Minimal Separating Trans.]
	Generate a minimal $(\Tin,\Tout)$-separating transaction for $\D$, or determine that one does not exist.
\end{problem}

\begin{problem}[Bounded Separating Trans.]
For a given number $k$, generate a $k$-bounded $(\Tin,\Tout)$-separating transaction for $\D$, or determine that one does not exist.
\end{problem}
	
The complexity of generating separating	transactions depends on the types of integrity constraints in $\ic$.  Given a set  $\Delta\subseteq\set{key,fd,ind}$, we use  $\MSEP(\Delta)$ and  $\KSEP(\Delta)$ to denote the minimal separating transaction problem, and the bounded separating transaction problem, respectively, for the given class of integrity constraints. Formally, $\MSEP(\Delta)$ (resp.\ $\KSEP(\Delta)$) is in complexity class $\mathcal C$ if, for all blockchain databases  $\D= (\R,\ic,\trans)$ allowed by $\Delta$, and for all  $\Tin,\Tout\subseteq \trans$, it is possible to generate a  minimal (resp.\ $k$-bounded) $(\Tin,\Tout)$-separating transaction for $\D$, or determine that one does not exist, in time~$\mathcal C$. 


%
%

\myheadingnl{Key and Functional Dependencies}
We start by considering $\MSEP(\Delta)$ and $\KSEP(\Delta)$ when $\Delta$ contains key constraints and functional dependencies. We will show that $\MSEP(\Delta)$ is in PTIME, while $\KSEP(\Delta)$ is NP-complete. 

\eat{
\begin{figure}[t]
\begin{algorithm}{\fdalgname}[(\R,\ic,\trans),\Tin,\Tout]{}
$\R_* \gets \R\cup_{T_i\in \Tin} T_i$ \\
\qif $\R_* \not\models \ic$ \\
\qthen return ``failed''	\qfi \\
$T\gets \emptyset$ \\
\qfor each $T_o\in \Tout$ \\
\qdo \qif $\R \cup T \cup T_o \models \ic$ \\
		\qthen $t\gets\qproc{Contradict}(T_o,\R_*\cup T,\ic)$ \\
					\qif $t = \bot$ \\
					\qthen return  ``failed'' \\
					\qelse $T\gets T\cup \set{t}$ \qfi \qfi \qrof\\

\qfor each $t\in T$ \\
\qif $T-\set{t}$ is  $(\Tin,\Tout)$-separating   \\
\qthen $T \gets T-\set{t}$ \qfi\qrof \\
return $T$
 \end{algorithm} 
\begin{algorithm}{Contradict}[T_o,\R,\ic]{}
		\qfor each $t_o\in T_o$ and $rel(t_o):X\rightarrow Y\in \ic$ \\
					\qdo $R \gets \set{t'\in \R \mid rel(t') = rel(t_o)}$ \\
					$t \gets \qproc{ChaseFD}(t_o,X,\ic,R)$\\
								\qif $\set{t_o,t}\not \models \ic$ \\
								\qthen 	return $t$ \qfi \qrof\\
return $\bot$ 
\end{algorithm}
\caption{An algorithm that finds a minimal separating transaction, for databases with key constraints and functional dependencies.\label{fig:msep:fd:key:algorithm}}
\end{figure}
}
\begin{figure}[t]
	\begin{algorithm}{\fdalgname}[(\R,\ic,\trans),\Tin,\Tout]{}
		$\R_* \gets \R\cup_{T_i\in \Tin} T_i$ \\
		\qif $\R_* \not\models \ic$ {\bf then} return ``failed''	\qfi \\
		$T\gets \emptyset$ \\
		\qfor each $T_o\in \Tout$ \\
		\qdo \qif $\R \cup T \cup T_o \models \ic$ \\
		\qthen $t\gets\qproc{Contradict}(T_o,\R_*\cup T,\ic)$ \\
		\qif $t = \bot$ \\
		\qthen return  ``failed'' \\
		\qelse $T\gets T\cup \set{t}$ \qfi \qfi \qrof\\
		
		\qfor each $t\in T$ \\
		\qif $T-\set{t}$ is  $(\Tin,\Tout)$-separating  {\bf then} $T \gets T-\set{t}$ \qfi\qrof \\
		return $T$
	\end{algorithm} 
	\begin{algorithm}{Contradict}[T_o,\R,\ic]{}
		\qfor each $t_o\in T_o$ and $rel(t_o):X\rightarrow Y\in \ic$ \\
		\qdo $R \gets \set{t'\in \R \mid rel(t') = rel(t_o)}$ \\
		$t \gets \qproc{ChaseFD}(t_o,X,\ic,R)$\\
		\qif $\set{t_o,t}\not \models \ic$ {\bf then} 	return $t$ \qfi \qrof\\
		return $\bot$ 
	\end{algorithm}
	\caption{An algorithm that finds a minimal separating transaction, for databases with key constraints and functional dependencies.\label{fig:msep:fd:key:algorithm}}
\end{figure}

A polynomial time algorithm for  $\MSEP(\set{key, fd})$, called \fdalg,
appears in 
 Figure~\ref{fig:msep:fd:key:algorithm}.
In Lines~1--3 of \fdalg, we check if there exists a possible world containing all of $\Tin$. If not, by definition, there can be no $(\Tin,\Tout)$-separating transaction. Assuming otherwise, we attempt to build a $(\Tin,\Tout)$-separating transaction $T$ and start by initializing $T$ to the empty set (Line~4).

For each $T_o\in \Tout$, we check if $T$, together with $\R$, already contradicts $T_o$, given the integrity constraints $\ic$. If not (and thus, the condition of Line~6 is true), we attempt to generate a tuple, to add to $T$, that will cause $\R\cup T\cup T_o \not\models \ic$ to hold. To this end, the subprocedure $\qproc{Contradict}$ is run. (This subprocedure is explained later on.) If $\qproc{Contradict}$ returns $\bot$, no appropriate tuple can be found, and thus, there is no $(\Tin,\Tout)$-separating transaction (Line~9). Otherwise, the tuple found  is added to $T$, and we proceed to the next transaction in $\Tout$. 
Finally, in Lines~11--13, we minimize the set $T$, so that in Line~14, a minimal $(\Tin,\Tout)$-separating transaction is returned.

In the subprocedure $\qproc{Contradict}$ we iterate over all pairs $t_o\in T_o$ and $X\rightarrow Y\in \ic$, such that the dependency $X\rightarrow Y$ is over the relation to which $t_o$ belongs (Line~1). In Line~2, we set $R$ to contain all tuples in $\R$ (actually this includes all tuples in the original $\R$, in $\Tin$ and in $T$) that belong to the same relation as $t_o$. Our goal is to find a tuple $t$ such that {\em (1)} $\set{t}\cup R\models \ic$ and {\em (2)}  $\set{t_o,t}\not\models \ic$.

To this end, we apply the procedure $\qproc{ChaseFD}$ in Line~3. This procedure is not provided in pseudo-code as it is more easily described in text. The procedure \qproc{ChaseFD} starts by initializing a new tuple $t$ to be equal to $t_o$ on all attributes in $X$, while all other attributes are given new unused values.  Then, we perform a chase by repeatedly checking if there is a dependency $U\rightarrow V$ in $\ic$ such that $t$ is equal to some other tuple $t'$ in $R$ on attributes $U$. If this is the case, we update $t$ to be equal to $t'$ on attributes $V$. (Note that this update changes the values in $t$, not in $t'$.) This ensures that the result is consistent with $\ic$. We can then show the following result.


\begin{restatable}{theorem}{thmAlgCorrect}\label{theorem:gen:fd}
Given a blockchain database with key constraints and functional dependencies, and sets of transactions $\Tin, \Tout$,  algorithm \fdalg returns a $(\Tin,\Tout)$-separating transaction, if one exists, and runs in PTIME.
\end{restatable}

\newcommand{\thmAlgCorrectProof}{
\begin{proof}
The polynomial runtime is immediate from examination of the algorithm. Hence, we focus on correctness. We first consider the case in which the algorithm returns a set $T$ (i.e., does not return ``failed''). The set $T$ is consistent with $\R$, as well as all of $\Tin$, as this is assured by applying the chase procedure in Line~3 of \qproc{Contradict}. It also holds that for all $T_o\in \Tout$, there is no possible world including both $T$ and $T_o$, as the inclusion of the tuple returned from \qproc{Contradict},  when called with $T_o$, contradicts $T_o$ given the integrity constraints (Line~4 of \qproc{Contradict}). Hence $T$ is $(\Tin,\Tout)$-separating. 

To see that $T$ is also minimal, observe that the following property is easy to show for $(\Tin,\Tout)$-separating transactions, when all integrity constraints are key constraints or functional dependencies:
Let $\set{t}\subseteq T_1\subseteq T_2$ be transactions such that both $T_1$ and $T_2$ are $(\Tin,\Tout)$-separating. Then, if $T_1-\set{t}$ is  $(\Tin,\Tout)$-separating, it also holds that $T_2-\set{t}$ is $(\Tin,\Tout)$-separating. This property implies that if we choose not to remove a tuple $t$ in the loop of Line~11 of \fdalg, it will not be possible to remove $t$ later on. Thus, the transaction returned by this function is indeed minimal. 

We now consider the cases in which   \fdalg returns ``failed''.
If  \fdalg returns ``failed'' on Line~3, there is no $(\Tin,\Tout)$-separating transaction, as $\R_*\not\models\ic$, and hence, there is no possible world containing $\R,\Tin, T$, for any $T$.
We show that if \fdalg returns ``failed'' on Line~9, then there is no $(\Tin,\Tout)$-separating transaction. 

If \fdalg returns ``failed'' upon the first iteration of the loop of Line~5, then clearly there is no $(\Tin,\Tout)$-separating transaction, as it is not possible to find a tuple that together with $T_o$ contradicts the integrity constraints,  while being consistent with $\R$ and $\Tin$. On subsequent iterations of the loop of Line~5, a return of ``failed'' implies that it is not possible to find a tuple $t$ that together with $T_o$ contradicts the integrity constraints, while being consistent with $\R$, $\Tin$ and the set $T$ created thus far. We now show that if this is the case, then it is not possible to find a tuple $t$ that together with $T_o$ contradicts the integrity constraints, while being consistent with $\R$ and $\Tin$ (i.e., our choice of $T$ is irrelevant to the ability to contradiction $T_o$). 

By way of contradiction, suppose that we return ``failed'' for transaction $T_o$, but there exists a tuple $t$ such that $\R_* \cup \set{t}\models \ic$ and $T_o\cup\set{t}\not\models\ic$. Then, there is a tuple $t_o\in T_o$ and an integrity constraint $X\rightarrow A$ over the relation of $t_o$, such that $t[X] = t_o[X]$, but $t[A] \neq t_o[A]$. (To simplify the proof, we assume, without loss of generality, that all functional dependencies have a single attribute on their righthand-side.) Clearly such a tuple can be derived by equating $t$ with $t_o$ on $X$, choosing new constants for all other attributes of $t$, and chasing the tuple $t$ with $\R_*$. Let $t_1$ be the result of chasing $t$ with~$\R_*$. 

Since we assume that we returned ``failed'' for $T_o$, it follows that the chaise of $t$ with $\R_* \cup T$ equated $t[A]$ to the same value as $t_o[A]$, even though, by assumption, the chase of $t$ with $\R_*$ did not do so. Let $t_2$ be the result of chasing $t$ with $\R_* \cup T$. It follows that there is some tuple $t'\in T$ and some $Y\rightarrow A$ in $\ic$ such that $t_2[Y] = t'[Y]$ and $t'[A] = t_o[A]$. 

Since $t_2[Y] = t'[Y]$, it follows that $Y$ is not a new constant. Therefore, there is some other tuple $t''\in \R_*$ for which $t''[Y] = t'[Y] = t_2[Y]$. Due to the functional dependency $Y\rightarrow A$, it follows that $t''[A] = t'[A] = t_o[A]$. It then follows that the chase of $t$ with $\R_*$ must have equated $t[A]$ to $t_o[A]$ (using $t''$), in contradiction to the assumption. Hence, it follows that if we return ``failed'' for transaction $T_o$,  there indeed does not exist a tuple $t$ such that $\R_* \cup \set{t}\models \ic$ and $T_o\cup\set{t}\not\models\ic$.
\end{proof}
}

While the above theorem shows that minimal separating transactions can be found in polynomial time, the next result shows that finding bounded-size  separating transactions  is hard. We note that the following result is of interest only when $k<|\Tout|$, as any minimal separating transaction will be of size at most $|\Tout|$ (containing one tuple at most for each $T_o\in \Tout$), and such minimal separating transactions are easily found. 

\begin{restatable}{theorem}{thmKSEPkey}
$\KSEP(\set{key})$ is NP-complete.
\end{restatable}

\newcommand{\thmKSEPkeyProof}{
\begin{proof}
Membership in NP can easily be shown, as a polynomial-sized witness can be guessed and verified. To see that a witness is  polynomial-sized, note that it if there is a witness, then there is one containing at most $\min(\set{|\Tout|,k})$ tuples. Note also that it is sufficient to guess constants from among a domain containing all constants in $\R, \Tin, \Tout$, as well as $|\Tout|$ arbitrary new constants for each attribute of each relation.

NP-hardness is shown by a reduction to hitting set. Let $U$ be a universe of values and let $\S$ be a set of sets of values from $U$. Given a value $k$, the hitting set problem is to find a subset $U'\subseteq U$ of size at most $k$, such that for each $S\in \S$, it holds that $U'\cap S\neq \emptyset$. 

Given an instance $U,\S,k$ of the hitting set problem, we create an instance of $\KSEP(\set{key},k)$ as follows. We assume there is a single relation $R(A,B)$. For each $S\in \S$, we create a transaction $T_S$ containing the tuples $\set{(x,0)\mid x\in S}$. Let $\D$ be the blockchain database $(\emptyset, \set{R:A\rightarrow B}, \trans)$ where $\trans = \set{T_S\mid S\in \S})$. Choose $\Tin = \emptyset$ and $\Tout = \trans$. 

Clearly,  there is a one-to-one correspondence between hitting sets of size $k$ and $(\Tin,\Tout)$-separating transactions of size $k$. Specifically, $x_1,\ldots,x_k$ is a hitting set if and only if $\set{(x_1,1),\ldots,(x_k,1)}$ is a $(\Tin,\Tout)$-separating transactions of size $k$.
\end{proof}
}

\myheading{Inclusion Dependencies}
We show that a minimal $(\Tin, \Tout)$-separating transaction can be found in polynomial time, if all integrity constraints are inclusion dependencies. 

\begin{restatable}{theorem}{thmGenInc}
	Given a blockchain database with inclusion dependencies, and sets of transactions $\Tin, \Tout$, 
it is possible to find a $(\Tin,\Tout)$-separating transaction, if one exists, in PTIME.
\end{restatable}

\newcommand{\thmGenIncProof}{
\begin{proof}
In order to find a $(\Tin,\Tout)$-separating transaction, we start by initializing $T$ to be the empty set. Let $c$ be a new unused constant.  Then, we iterate over the tuples $t_i$ in the transactions of $\Tin$. While there is some inclusion dependency $R[X_1,\ldots,X_n]\subseteq S[Y_1,\ldots,Y_n]$ over the relation $R$ of $t_i$, such that there is no tuple $t'$ in $S$ with $t_i[X_1,\ldots,X_n] = t'[Y_1,\ldots,Y_n]$, we create such a tuple by generating a tuple $t$ over $S$ with  
\begin{itemize}
\item $t[Y_1,\ldots,Y_n] = t_i[X_1,\ldots,X_n]$ and
\item the constant $c$ for all other attributes.
\end{itemize}
We then add $t$ to $T$. We continue chasing the newly generated tuple $t$ with respect to the integrity constraints, and adding new tuples as required. In all undetermined attributes, we set the value to the same constant $c$. 

Observe that the above process is finitely terminating (and polynomial, in fact), since all ``new'' values are set to $c$. Observe too, that at the end of this process, it is guaranteed that 
\[ \R \cup_{T_i\in \Tin}T_i \cup T \models \ic\,.\]
However, it is possible that $T$ is not $(\Tin,\Tout)$-separating as it may not meet the requirements for $\Tout$. 

In order to determine if $T$ is $(\Tin,\Tout)$-separating we find the single maximal possible world, by starting with $\R$ and then adding transactions from $\trans\cup{T}$ that are consistent with the integrity constraints. If this process allows us to add a transaction from $\Tout$, we conclude that $T$ is not a $(\Tin,\Tout)$-separating transaction. Indeed, we can actually conclude that there is no $(\Tin,\Tout)$-separating transaction, as the above procedure is simply a minimal chasing of $\Tin$ with $\ic$, and our choice of values for undefined constants was always the same value $c$, which does not, in itself, aid in satisfying the integrity constraints for tuples in $\Tout$. 

It now remains to minimize $T$. This is accomplished as in Lines~11-13 of \fdalg, by iteratively attempting to remove a tuple $t$, and checking if the result is still a $(\Tin,\Tout)$-separating transaction. Note that the latter is accomplished by iteratively attempting to expand $\R$ by transactions in $\trans$ and other tuples in $T$, until all of $\Tin$ can be included---in which case $t$ is not needed, or until termination---in which case $t$ is necessary.  The result is guaranteed to be minimal for the same reasons as in Theorem~\ref{theorem:gen:fd}. 
\end{proof}
}

Finding a $k$-bounded $(\Tin,\Tout)$-separating transaction is more difficult. Indeed this problem is NP-complete, even if all inclusion dependencies are acyclic. 
For completeness, we recall the definition of acyclic inclusion dependencies. Let $\ic$ be a set of inclusion dependencies over relations $R_1,\ldots,R_n$. The {\em dependency graph\/} for $\ic$ contains a node for each relation $R_i$, and an edge from $R_i$ to $R_j$ if there is a inclusion dependency $R_i[X_1,\ldots,X_n]\subseteq R_j[Y_1,\ldots,Y_n]\in \ic$. 
A set of inclusion dependencies is {\em acyclic\/} if its dependency graph is acyclic. 

We can show the following result. 
\begin{restatable}{theorem}{thmKSEPind}
$\KSEP(\set{ind})$ is NP-complete, even if all inclusion dependencies are acyclic.
\end{restatable}

\newcommand{\thmKSEPindProof}{
\begin{proof}
We show the required by a reduction from 3-SAT.  Let $\phi_1\wedge \cdots \wedge \phi_n$ be a 3-SAT formula, where $\phi_i$ is a disjunction of variables or their negation from $x_1,\ldots,x_k$. We create a blockchain database, in which the set of relations is: Assign1$(X,V)$, Assign2$(X,V)$, Truth$(V)$, Sat$(X_1,\ldots,X_k)$, Clause1$(C)$, Clause2$(C)$. The relation Truth contains two rows $(t),(f)$, while all other relations are empty. In addition, we have the inclusion dependencies
\begin{align*}
\text{Assign1}[V]&\subseteq \text{Truth}[V] \\
\text{Assign2}[X,V]&\subseteq \text{Assign1}[X,V] \\
(\forall i\leq k)\,\, \text{Sat}[X_i]&\subseteq\text{Assign1}[X]\\
\text{Clause2}[C]&\subseteq \text{Clause1}[C] 
\end{align*}
The set $\trans$ includes the following transactions:
\begin{itemize}
\item for each $i\leq k$, there is a transaction containing the tuples Assign2$(x_i,t)$, as well as Clause1$(c_j)$, for all $j$ such that $x_i$ appears positively in clause $j$;
\item for each $i\leq k$, there is a transaction containing the tuples Assign2$(x_i,f)$, as well as Clause1$(c_j)$, for all $j$ such that $x_i$ appears negatively in clause $j$;
\item a transaction containing $\text{Clause2}(c_1), \ldots, \text{Clause2}(c_n),$ $\text{Sat}(x_1,\ldots,x_k)$. 
\end{itemize}
The last of the transactions described above is the only transaction in $\Tin$. The set $\Tout$ is empty. 

We now show that there is a $(\Tin,\Tout)$-separating transaction of size $k$ if and only if there is a satisfying assignment for the 3-SAT formula. For the first direction, suppose there is a satisfying assignment for the 3-SAT formula. We define the transaction $T$ to contains $\text{Assign1}(x,t)$ for each variable $x$ assigned to true, and $\text{Assign1}(x,f)$ for all variables assigned to false. Now, observe that there is a possible world containing $\Tin$, i.e., precisely the world containing $T$, along with the transactions containing $\text{Assign2}(x,t)$ and $\text{Assign2}(x,f)$, for all variables assigned true and false, respectively. 

Conversely, suppose that there is a $(\Tin,\Tout)$-separating transaction $T$ of size $k$. In order for there to be a possible world including $\Tin$, the transaction $T$ must contain an atom of the form  $\text{Assign1}(x,t)$  or  $\text{Assign1}(x,f)$, for each variable $x$ (so that the inclusion dependency on Sat will be satisfied). Since $T$ is of size at most $k$, there must be precisely one item of this type for each variable $x$. Now, in order for $\Tin$ to be part of a possible world, we must be able to satisfy the inclusion dependency of Clause2. This will occur if and only if the atoms in $T$ form a truth assignment in the 3-SAT formula. 
\end{proof}
}

\myheading{Functional and Inclusion Dependencies}
We now consider the setting in which $\ic$ may include key constraints, functional dependencies and inclusion dependencies. We show that determining whether there exists a $(\Tin,\Tout)$-separating transaction is undecidable in this case. Undecidability follows from the fact that the implication problem for functional and inclusion dependencies is undecidable.

\begin{restatable}{theorem}{thmUndecidable}
The problem of determining whether there exists a $(\Tin,\Tout)$-separating transaction, when $\ic$ can include functional dependencies and inclusion dependencies, is undecidable. 
\end{restatable}

\newcommand{\thmUndecidableProof}{
\begin{proof}
We show the required by a reduction from the implication problem. In~\cite{DBLP:journals/siamcomp/ChandraV85}, it was shown that, given a set of functional dependencies $F$ and a set of inclusion dependencies $I$, as well as a functional (or inclusion) dependency $f$, the problem of determining whether $F\cup T\models f$ is undecidable. Given an instance of this problem, i.e., $F$, $I$, and a functional dependency $f=R:X_1,\ldots,X_n\rightarrow A$, we create an instance of the $(\Tin,\Tout)$-separating transaction problem as follows. 
\begin{itemize}
\item $\R$ contains empty instances of all relations mentioned in $F$ or $I$, as well as a relation $S$ with schema $(X_1,\ldots,X_n,A)$;
\item $\ic$ contains $F$, $I$ and the inclusion dependency \[S[X_1,\ldots,X_n,A]\subseteq R[X_1,\ldots,X_n,A]\,;\]
\item $\trans=\Tin$ contains the single transaction \[\set{S(x_1,\ldots,x_n,a_1),S(x_1,\ldots,x_n,a_2)}\,.\]
\end{itemize}
Now, there is a $(\Tin,\Tout)$-separating transaction if and only if there is some instance of the relations in which $R$ contradicts the functional dependency $f$, as it must include the values of the tuples in $S$. Thus, there is a $(\Tin,\Tout)$-separating transaction if and only if $F\cup T\not\models f$, thereby showing undecidability, as required. 
\end{proof}
}

\section{Future Research and Open Problems} \label{sec:open}
Our work has only taken initial steps in answering questions related to \replaced{databases based on blockchains.}{ blockchain databases.} In this section we list open questions and possible extensions of the research we have presented. 

\myheadingnl{Economic Considerations for Possible Worlds}
When issuing a transaction, the user defines a fee that will be accrued by the miner that incorporates the transaction into a block.
Miners choose transactions in a way that will maximize the fees accumulated. Due to the dependencies between transactions, the strategies  choosing which transactions to include are complex. 
In our model, all possible worlds are equally valid. However, when taking the economic realities into consideration, this implies some type of weighting or probabilities to the possible worlds. Interesting questions, such as the contradiction of denial constraints in likely worlds, or generating likely transactions, then arise. 

\myheadingnl{Generating Transactions with Constraints}
In this paper, we studied the problem of generating $(\Tin,\Tout)$-separating transactions. Indeed, automatic transaction generation is an important tool to aid the user in formulating transactions that avoid undesirable outcomes. In the setting considered thus far, we did not consider additional user-defined constraints on the transaction being generated, and hence, e.g., could choose any new constants to be used in the transactions generated. An important generalization of this work is to allow the user to specify constraints (perhaps in the form of a query or as domain information) that the transaction generated must satisfy. Devising efficient algorithms for transaction generation in this setting is both useful and challenging.

\deleted{
\myheadingnl{Sharding blockchains} 
By design, blockchain databases are extremely space-inefficient, as they typically replicate all data at each node. 
While there is a great deal of work on sharding database systems, the problem of sharding permissionless blockchain databases remains open. It is unclear how to maintain consistency and consensus in an open system in which different entities hold copies of different shards of the database. Specifically, if a malicious entity manages to have the only copy of a certain shard of the database, it can render the information in that piece unavailable to others, thereby erasing critical data (e.g., financial records) that cannot be otherwise recovered. 
}
\myheadingnl{Database management under \deleted{more }advanced protocols such as the Lightning network}
Blockchain databases typically suffer from low throughput and high transaction acceptance time. Off-chain transaction channels, along with their composition into the Lightning Network~\cite{poon2015bitcoin}, make up one of the \deleted{main }proposals put forth to alleviate these issues. \deleted{Intuitively,} Off-chain transaction channels form a protocol layer that assumes the existence of blockchains. The basic elements of off-chain transaction channels are regular transactions that are exchanged between the participants in the protocol. These transactions aggregate several payments over time, and are admitted into the blockchain only at the closure of the channel, to finalize the aggregate result of the interactions. 
The transactions that are exchanged are created in a way as to avoid undesirable states, such as one participant stealing money using intermediate states. 

Just like our work considers possible worlds that the blockchain database may end up in, security of the Lighting Network protocol depends strongly on the same type of analysis. This can be seen as a version of our problem, but on steroids. Given its importance for scalability, this is a highly important problem.

\myheadingnl{Detecting independent or commutative transaction sets} Some protocols can offer improvements to transaction processing via the consensus mechanism if it can be guaranteed that transactions are independent of one another in some manner. The SPECTRE protocol~\cite{sompolinsky2016spectre} for example, can provide faster processing times for transactions as long as they can be committed to the database in any order (i.e., they are commutative). While this is not generally the case for arbitrary data, this does hold true for simple payments that come from different sources. The question of identifying such commutative transactions (or sets of transactions) and utilizing similar fast consensus mechanisms in databases, other than cryptocurrencies, is an important open problem. 

\deleted{
\myheadingnl{Considering more elaborate state spaces and general smart contracts} Blockchain systems such as Ethereum utilize the shared database to hold state for various smart-contracts that are essentially Turing complete computer programs. The question of possible states of the database in the presence of more sophisticated behaviors induced by such programs is an especially interesting extension of our current work. While the problem for general purpose computations may be difficult, there is sufficient interest in other more limited scripting systems that may describe state changes. More limited languages have been considered in attempt so as to ensure that there are fewer bugs in smart-contract code. In these languages, understanding possible worlds may be tractable.
}
\myheadingnl{More advanced strategies for Bitcoin unspent output management}
Bitcoin wallets effectively set aside several transaction outputs of different sizes that are to be used when composing new transactions. Once a transaction is prepared, the output used by that transaction cannot be reused. To maintain flexibility, wallet software prepares outputs of different sizes that can be composed together to provide payments. The management of such output, while taking into  consideration fees that are associated with producing the related transactions, can be greatly improved and will gain much from formal treatment. This problem is akin to maintaining a level of flexibility that will allow one to cheaply respond to requests from the user on unknown inputs that may come in the future. This is a special new type of transaction generation problem.


\section{Conclusion} \label{sec:conclusion}

This paper studied theoretical foundations for blockchain databases. To this end, we formally defined a blockchain database, and introduced two important problems. First, we considered the complexity of denial constraint satisfaction. The ability to determine if a denial constraint will always hold is critical in deciding if a new transaction can be safely issued. Second, we studied the problem of automatically generating transactions that can be used to separate between sets of pending transactions. This again, is pivotal in order to ensure that a new transaction cannot be included together with some already issued transactions. 

Blockchain databases are  a cutting-edge technology, and there are much efforts in the industry to bring blockchain databases to the developer and end user. These efforts are insufficient without thorough study of the ramifications of this new model on critical database operations. This paper has begun such a study and   charted open problems of interest that should be considered by the database community. As such, we hope this paper will serve as a springboard for active research in the important and timely area of blockchain databases.

\bibliography{main}

\appendix

\newcommand{\B}{\mathcal B}

\section{Integrity Constraints}
In this paper we consider key constraints, functional dependencies and inclusion dependencies. These notions are defined next. 
Let $R$ be a relation over attributes $A_1,\ldots,A_n$. Let $X$ and $Y$ be subsets of $\set{A_1,\ldots,A_n}$. A {\em functional dependency} has the form 
\[X \rightarrow Y\,.\]
 We say that this dependency is satisfied by $R$, if, for every two tuples $t$ and $s$ in $R$ it holds that 
	\[ \left(\bigwedge_{A\in X}t[A] = s[A]\right)\Rightarrow \left(\bigwedge_{B\in Y}t[B] = s[B\right)\,.\]
A {\em key constraint\/} is simply a functional dependency in which $Y = \set{A_1,\ldots,A_n}$. 

Finally, let $R'$ be a relation over attributes $B_1,\ldots,B_m$.
An {\em inclusion dependency\/} has the form 
\[ R[A_{i_1},\ldots A_{i_k}]\subseteq R'[B_{j_1},\ldots,B_{j_k}]\]
and is satisfied by $R$ and $R'$ if, for every tuple $t\in R$, there exists a tuple $t'\in R'$ such that 
\[ \bigwedge_{l\leq k}]t[A_{i_l}] = t'[B_{j_l}] \,.\]

\section{Aggregate Functions}
In this paper, we consider four aggregate functions, namely, $\COUNT$, $\COUNTD$ (count distinct), $\SUM$ and $\MAX$. The functions $\COUNT$ and $\COUNTD$ are well-defined when applied to a bag containing tuples of any fixed arity, while $\SUM$ and $\MAX$ are well defined only when applied to bags containing tuples of arity one. Moreover, $\SUM$ is applicable only over numerical domains, and $\MAX$ is applicable over domains that are completely ordered. 

Given a bag $\B$, let $\mathit{vals}(\B)$ be the set of values appearing in $\B$, and let $\mathit{mult}(v,\B)$ be the multiplicity of $v$ in $\B$. 
The semantics of these functions are formally defined as follows for a bag $\B$:
\begin{align*}
	\COUNT(\B) &= \sum_{v\in \mathit{vals}(\B)}\mathit{mult}(v,\B)  & \SUM(\B) &= \sum_{v\in \mathit{vals}(\B)}  \mathit{mult}(v,\B)\times v\\
	\COUNTD(\B) &=  |\mathit{vals}(\B)|   & \MAX(\B) &= \max\set{v\mid v\in \mathit{vals}(\B)}\,.
	\end{align*}

\section{Proofs}

\propPTIME*
\propPTIMEProof

\thmComplexityConjunctive*
\thmComplexityConjunctiveProof

\begin{table}
	\begin{center}
		
		\begin{tabular}{lllll} \toprule
			Aggregate function & Comparison & Integrity Constraints & Complexity  \\\midrule
			$\MAX$ & $>, <, =$ & key, fd & PTIME  \\
			& $>$ & ind & PTIME  \\
			& $<, =$ & ind & CoNP-C \\ 
			& $>, <, =$ & key, ind & CoNP-C \\ 
			$\COUNT$, $\COUNTD$, $\SUM$ & $<$ & key, fd & PTIME \\
			& $>, =$ & key & CoNP-C \\
			& $>$ & ind & PTIME (positive queries) \\
			& & & 		 CoNP-C (queries with negation) \\
			& $<,=$ & ind & CoNP-C \\
		\end{tabular}
	\end{center}
	\caption{Summary of complexity results for the denial constraint satisfaction problem for positive query classes with aggregation. Since key constraints are a special case of functional dependencies, all hardness results for keys still hold in the presence of functional dependencies.\label{table:complexity}}
\end{table}
\thmComplexityAgg*
\thmComplexityAggProof

\thmAlgCorrect*

\thmAlgCorrectProof

\thmKSEPkey*

\thmKSEPkeyProof

\thmGenInc*

\thmGenIncProof

\thmKSEPind*

\thmKSEPindProof

\thmUndecidable*

\thmUndecidableProof

\end{document}